\documentclass[fleqn,twoside,a4paper]{article}
\usepackage[headings]{espcrc2}
\usepackage{amsmath,epsfig}
\usepackage{psfrag}

\newcommand{\cha}{\tilde{\chi}}
\newcommand{\neu}{\tilde{\chi}^0}
\newcommand{\sn}{{\tilde{\nu}}}

\newcommand{\snl}{{\tilde{\nu}_l}}

\newcommand{\gev}{\,\, \mathrm{GeV}}

\newcommand{\RR}{{\rm R}}
\newcommand{\LL}{{\rm L}}

\title{Reconstructing Supersymmetry at ILC/LHC}

\author{
G.~A.~Blair\address[DESY]{Deutsches Elektronen-Synchrotron DESY,
                          D-22603 Hamburg, Germany}
           \address{Royal Holloway University of London, Egham, Surrey. TW20 0EX,
                    UK},
A.~Freitas\address{Institut f\"ur Theoretische Physik, Universit\"at Z\"urich, 
		CH-8057 Z\"urich, Switzerland},
H.-U.~Martyn\address{I. Physik. Institut, RWTH Aachen, D-52074 Aachen, Germany},
G.~Polesello\address[CERN]{CERN, Department of Physics, CH-1211 Geneva 23,
                                                                Switzerland},
W.~Porod\address{IFIC - Instituto de F\'\i sica Corpuscular,
                 E-46071 Valencia, Spain},
and P.~M.~Zerwas\addressmark[DESY]
} 

\begin{document}

\begin{abstract}
\vspace{-1.8cm}
\ \hfill {\parbox[b]{2cm}{\small\flushright DESY
05--240\\IFIC/05--64\\ZH--TH 25/05}}\\

Coherent analyses of experimental results from LHC and ILC
will allow us to draw a comprehensive and precise picture of the
supersymmetric particle sector. Based on this platform the fundamental
supersymmetric theory can be reconstructed at the high scale which is potentially
close to the Planck scale. This procedure will be reviewed for three
characteristic examples: minimal supergravity as the paradigm;
a left-right symmetric extension incorporating intermediate mass scales;
and a specific realization of string effective theories.
\end{abstract}

\maketitle

\section{Introduction}

It is widely accepted that particle physics is rooted at the Planck
scale where it is
intimately linked with gravity. A stable bridge between the two vastly
different scales, the electroweak scale of order 100 GeV and
the grand unification / Planck scale of order $10^{16}$ / $10^{19}$
GeV, is built by supersymmetry which is characterized by a typical scale of
order TeV. If this picture is realized in nature, methods must be
developed which allow us to reconstruct the fundamental physics scenario
near the grand unification / Planck scale. Elements of the picture can be
provided by experiments observing proton decay, neutrino physics within the
seesaw frame, various aspects of cosmology and, last not
least, high-precision experiments at high energies \cite{R1}.

The reconstruction of a physical scenario more than fourteen orders of
magnitude above accelerator energies is a demanding task. Nevertheless,
the extrapolation of the gauge couplings to the grand unification scale
by renormalization group methods is an encouraging example
\cite{R2}. In
supersymmetric theories a rich ensemble of soft breaking parameters can
be investigated. Symmetries and the impact of high-scale parameters
can be studied which will reveal essential elements of the fundamental
physics scenario \cite{R3}.

\vspace{1.5ex}
The proton-collider LHC and the $e^+e^-$ linear collider ILC,
now in the design phase, are a perfect tandem for exploring supersymmetry.
The heavy colored supersymmetric particles, squarks and gluinos, can be
generated for masses up to 3 TeV with large rates at LHC. Subsequent
cascade decays give access to lower mass particles \cite{R4}. The
properties of
the potentially lighter non-colored particles, charginos/neutralinos and
sleptons, can be studied very precisely at an $e^+e^-$ linear collider
\cite{R5} by exploiting in particular polarization phenomena at
such a lepton facility \cite{R5A}. After the properties of the light particles are
determined precisely, the properties of the heavier particles can
subsequently be studied in the cascade decays with similar precision.
Coherent LHC and ILC analyses \cite{R1} will thus provide us with a
comprehensive
and high-precision picture of supersymmetry at the electroweak scale,
defining a solid platform for the reconstruction of the fundamental
supersymmetric theory near the Planck scale.

This procedure will be described for three characteristic examples in
this report -- minimal supergravity, a left-right symmetric extension,
and a string effective theory.

\vspace{1.5ex} 
(a) {\underline{Minimal supergravity [mSUGRA]}} \cite{R6} defines a
scenario in which these general ideas can be quantified most easily
due to the rather simple structure of the theory. Supersymmetry
is broken in a hidden sector and the breaking is transmitted to our
eigen-world by gravity. This suggests the universality of the soft SUSY
breaking parameters -- gaugino / scalar masses and trilinear couplings --
at the high scale which is generally identified with the grand unification
scale of the gauge couplings.

\vspace{1.5ex} 
(b) {\underline{Left-right symmetric extension:}} If
in LR supersymmetric theories \cite{R7} non-zero neutrino
masses are introduced by means of the seesaw mechanism \cite{R8},
intermediate scales
associated with the righ-handed neutrino masses of order $10^{10}$ to
$10^{14}$ GeV affect the evolution of the supersymmetry parameters from
the electroweak to the unification scale \cite{R9,R3}. If combined
with universality
at the unification scale, the intermediate scale modifies the observable
scalar mass parameters in the lepton sector of the third generation
at the electroweak scale. This effect can be exploited to estimate
the seesaw scale \cite{R10}, thus giving indirectly experimental access
to the fundamental high-scale parameter in the neutrino sector.

\vspace{1.5ex} 
(c) {\underline{String effective theory:}} In orbifold compactifications
of heterotic string theories, the universality of the scalar mass
parameters is broken if these masses are generated by interactions
with moduli fields \cite{R11}. Since the interactions are determined by
modular weights that are integer numbers, the pattern of scalar masses is
characteristic for such a scenario \cite{R3}. Thus this type of string
theories can be tested stringently by measuring the modular weights.

\section{Minimal Supergravity}

\begin{figure}
\epsfig{figure=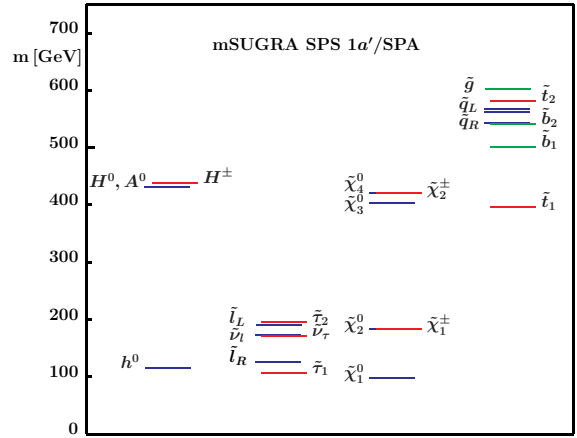, width=7.5cm}
\vspace{-3em}
\caption{Supersymmetric mass spectrum for SPS1a' scenario.}
\label{fig:SPS1a}
\end{figure}

The mSUGRA reference point SPS1a$'$ \cite{R12}, a derivative of the
Snowmass point
SPS1a \cite{R13}, is characterized by the following values of the soft
parameters
at the grand unfication scale:
\begin{eqnarray}
  \begin{array}{ll}
    M_{1/2} = 250~{\rm GeV} \qquad & M_0=70~{\rm GeV} \\
    A_0=-300~{\rm GeV} & {\rm sign}(\mu)=+\\
    \tan\beta=10 & \\
  \end{array}
\label{eq:sps1a}
\end{eqnarray}
The universal gaugino mass is denoted by $M_{1/2}$, the scalar mass by
$M_0$ and the trilinear coupling by $A_0$; the sign of the higgsino mass
parameter is chosen positive and $\tan\beta$, the ratio of the
vacuum-expectation values of the two Higgs fields, in the medium range.
The modulus of the higgsino mass parameter is fixed by requiring
radiative electroweak symmetry breaking so that finally $\mu = +396$ GeV.
The form of the supersymmetric mass spectrum in SPS1a$'$ is shown
in Fig.~\ref{fig:SPS1a}.  In this scenario the squarks and
gluinos can be studied very well at the LHC while the non-colored
gauginos and sleptons can be analyzed partly at LHC and in comprehensive
and precise form at an $e^+$e$^-$ linear collider operating at a total
energy up to 1 TeV with high integrated luminosity of 1~ab$^{-1}$.

\vspace{1.5ex} 
At {\underline{\bf{LHC}}} the masses can best be obtained by analyzing
edge effects in the cascade decay spectra \cite{R14}. The basic
starting point is the identification of a sequence of two-body decays:
\mbox{$\tilde q_L\rightarrow\tilde\chi^0_2 q\rightarrow\tilde\ell_R\ell q
\rightarrow \tilde\chi^0_1\ell\ell q$}.
The kinematic edges and thresholds predicted in
the invariant mass distributions of the two leptons and the jet
determine the masses in a model-independent way.
The four sparticle masses [$\tilde q_L$, $\tilde\chi^0_2$,
$\tilde\ell_R$ and $\tilde\chi^0_1$] are used subsequently as input for
additional decay chains like
\mbox{$\tilde g\rightarrow\tilde b_1 b\rightarrow \tilde\chi^0_2 bb$},
and the shorter chains \mbox{$\tilde q_R\rightarrow q \tilde\chi^0_1$}
and \mbox{$\tilde\chi^0_4\rightarrow\tilde\ell\ell$},
which all require the
knowledge of the sparticle masses downstream of the cascades. Residual
ambiguities and the strong correlations between the heavier masses and the
LSP mass are resolved by adding the results from ILC measurements which
improve the picture significantly.

\begin{figure}
\epsfig{figure=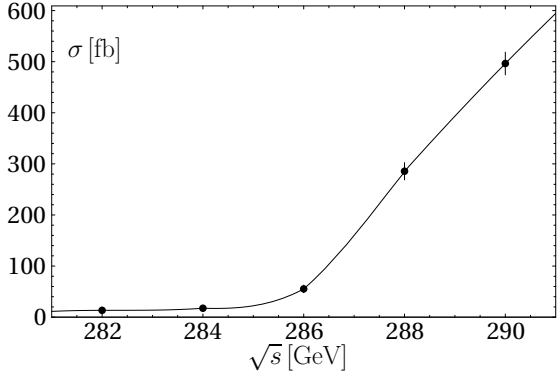, width=7.5cm, bb=30 390 472 682}
\vspace{-3em}
\caption{Excitation curve for $e^-e^-\to \tilde e_R^- \tilde e_R^-$ near
threshold, including background, initial-state radiation, beamstrahlung and
Couloumb correction. The cross-section rises steeply in S-wave production.
Ref.~\cite{R16}.}
\label{fig:thr}
\end{figure}
\begin{figure}
\epsfig{figure=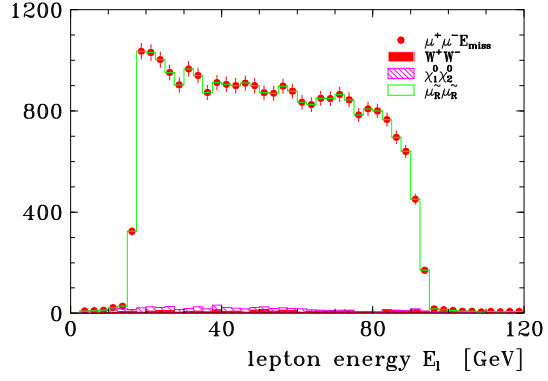, width=7.5cm, bb=70 228 500 520}
\vspace{-3em}
\caption{Energy sprectum of the muon from the decay ${\tilde{\mu}}^-_R \to \mu^-
{\tilde{\chi}}^0_1$, with important backgrounds and including initial-state
radiation and beamstrahlung effects. Ref.~\cite{R17}.}
\label{fig:decay}
\end{figure}

\vspace{1.5ex} 
At {\underline{\bf{ILC}}} very precise mass values can be extracted
from threshold scans and decay spectra \cite{R15}.
The excitation curves for chargino production
in S-waves rise steeply with the velocity of the particles
near the thresholds and they are thus very sensitive to their mass
values; the same holds true for mixed-chiral selectron pairs
in $e^+e^-\to \tilde e_R^+ \tilde e_L^-$
and for diagonal pairs in
$e^-e^-\to \tilde e_R^- \tilde e_R^-, \;  \tilde e_L^- \tilde e_L^-$
collisions, cf. Fig.~\ref{fig:thr}. 
Other scalar sfermions, as well as neutralinos,
are produced generally in P-waves, with a
less steep threshold behavior proportional to the
third power of the velocity.  Additional information,
in particular on the lightest neutralino $\tilde{\chi}^0_1$, can
be obtained from the very sharp edges of 2-body decay spectra,
such as ${\tilde{l}}^-_R \to l^- {\tilde{\chi}}^0_1$, cf. Fig.~\ref{fig:decay}.%

\vspace{1.5ex} 
The values of typical mass parameters and their related measurement
errors are presented in Tab.~\ref{tab:massesA}:
``LHC'' from LHC analyses and ''ILC'' from ILC analyses.
The third column ``LHC+ILC'' presents the corresponding errors if the
experimental analyses are performed coherently, i.e. the
light particle spectrum, studied at ILC with high precision,
is used as input set for the LHC analysis.
\renewcommand{\arraystretch}{1.2}
\begin{table}[h] \footnotesize
\begin{center} $
  \begin{array}{|c|c||c|c||c|}
    \hline 
    \ \mbox{Particle} \ &
    \ \ \mbox{Mass}\ \  & \mbox{``LHC''} & \mbox{``ILC''} 
                        & \mbox{``LHC+ILC''}\\ 
    \hline\hline
    h^0                 & 116.9 & 0.25 & 0.05 & 0.05 \\
    H^0                 & 425.0 &      & 1.5  & 1.5  \\
    \hline 
    \tilde{\chi}^0_1    &  97.7 & 4.8  & 0.05 & 0.05 \\
    \tilde{\chi}^0_2    & 183.9 & 4.7  & 1.2  & 0.08 \\
    \tilde{\chi}^0_4    & 413.9 & 5.1  & 3-5  & 2.5  \\
    \tilde{\chi}^\pm_1  & 183.7 &      & 0.55 & 0.55 \\ \hline 
    \tilde{e}_R         & 125.3 & 4.8  & 0.05 & 0.05 \\
    \tilde{e}_L         & 189.9 & 5.0  & 0.18 & 0.18 \\
    \tilde{\tau}_1      & 107.9 & 5-8  & 0.24 & 0.24 \\ \hline
    \tilde{q}_R         & 547.2 & 7-12 & -    & 5-11 \\
    \tilde{q}_L         & 564.7 & 8.7  & -    & 4.9  \\
    \tilde{t}_1         & 366.5 &      & 1.9  & 1.9  \\
    \tilde{b}_1         & 506.3 & 7.5  & -    & 5.7  \\ \hline
    \tilde{g}           & 607.1 & 8.0  & -    & 6.5  \\ \hline
  \end{array}$ \\
\end{center}
\caption{Accuracies for representative mass measurements
    of SUSY particles in individual LHC, ILC and 
    coherent ``LHC+ILC'' analyses
    for the reference point SPS1a$'$ [masses in {\rm GeV}].
    $\tilde q_R$ and $\tilde q_L$ represent the flavors
    $q=u,d,c,s$. [Errors presently extrapolated from SPS1a simulations.] }
\label{tab:massesA}
\end{table}

Mixing parameters must be extracted from measurements of cross
sections and polarization asymmetries,
in particular from the production of chargino pairs and
neutralino pairs, both in diagonal or mixed form \cite{R18}:
$e^+e^- \rightarrow {\tilde{\chi}^+_i}{\tilde{\chi}^-_j}$
\mbox{[$i$,$j$ = 1,2]} and ${\tilde{\chi}^0_i} {\tilde{\chi}^0_j}$
[$i$,$j$ = 1,$\dots$,4].
The production cross sections for
charginos are binomials of $\cos\,2\phi_{L,R}$, the mixing angles
rotating current to mass eigenstates. Using polarized electron
and positron beams, the mixings can be determined in a model-independent
way.

\vspace{1.5ex}
The fundamental SUSY parameters can be derived to lowest order
in analytic form \cite{R18}:
\begin{align}
\left|\mu\right|&=M_W[\Sigma + \Delta[\cos2\phi_R+\cos2\phi_L]]^{1/2}
\nonumber\\ \pagebreak
M_2&=M_W[\Sigma - \Delta(\cos2\phi_R+\cos2\phi_L)]^{1/2}\nonumber
\displaybreak[2] \\ 
|M_1|&= \left[ \textstyle \sum_i m^2_{\tilde{\chi}_i^0}
                 -M^2_2-\mu^2-2M^2_Z\right]^{1/2}
\nonumber\displaybreak[2]\\
|M_3|&=m_{\tilde{g}} \nonumber\displaybreak[1]\\
\tan\beta&=\left[\frac{1+\Delta (\cos 2\phi_R-\cos 2\phi_L)}
           {1-\Delta (\cos 2\phi_R-\cos 2\phi_L)}\right]^{1/2}
\label{eqn:basicLE}
\end{align}
where $\Delta =
(m^2_{\tilde{\chi}^\pm_2}-m^2_{\tilde{\chi}^\pm_1})/(4M^2_W)$
and
$\Sigma =  (m^2_{\tilde{\chi}^\pm_2}+m^2_{\tilde{\chi}^\pm_1})/(2M^2_W)
-1$.
The signs of $\mu$, $M_{1,3}$ relative to $M_2$ follow
from similar relations
and from cross sections for ${\tilde{\chi}}$ production and $\tilde{g}$
processes. 

The mass parameters of
the sfermions are directly related to the
physical masses if mixing effects are negligible:
\begin{equation}
m^2_{\tilde{f}_{L,R}}=M^2_{L,R}+m^2_f + D_{L,R}
\end{equation}
with $D_{L} = (T_3 - e_f \sin^2 \theta_W) \cos 2 \beta \, m^2_Z$
and $D_{R} = e_f \sin^2 \theta_W \cos 2 \beta \, m^2_Z$
denoting the D-terms. The non-trivial
mixing angles in the sfermion sector of the third generation
follow from the sfermion production cross sections \cite{R19} for
longitudinally polarized e$^+$/e$^-$ beams, which are bilinear
in $\cos$/$\sin2\theta_{\tilde f}$. The mixing angles and the two
physical
sfermion masses are related to the tri-linear couplings $A_f$,
the higgsino mass parameter $\mu$ and $\tan\beta(\cot\beta)$
for down(up) type sfermions by:
\begin{equation}
A_f-\mu\tan\beta(\cot\beta)=
\frac{m^2_{\tilde{f}_1}-m^2_{\tilde{f}_2}}{2 m_f}\sin2\theta_{\tilde f}
\end{equation}
$A_f$ may be determined in the $\tilde{f}$ sector if $\mu$
has been measured in the chargino sector. This procedure can be
applied in the stop sector.
Heavy Higgs $H,A$ decays to stau pairs may be used to determine the $A$
parameter in the stau sector \cite{R20}.

Refined analysis programs have been developed which include one-loop corrections
in determining the Lagrangian parameters from masses and cross sections
\cite{R21,R22} (see also \cite{r22A}).

\vspace{1.5ex} 
These measurements define the initial values for the evolution of the
gauge couplings and the soft SUSY breaking parameters to the grand
unification scale. The values at the electroweak scale are connected
to the fundamental parameters at the GUT scale by the renormalization
group equations. To leading order, \\[1ex]
\noindent
\renewcommand{\arraystretch}{1.2}
\begin{tabular}{ll}
 gauge couplings &: $\alpha_i = Z_i \, \alpha_U$ \hfill (5) \\
 gaugino masses  &: $M_i = Z_i \, M_{1/2}$ \hfill (6) \\
 scalar masses   &:   \\
 \multicolumn{2}{c}{$\hspace*{1cm} M^2_{\tilde\jmath} = M^2_0 + c_j
M^2_{1/2} +
        \sum_{\beta=1}^2 c'_{j \beta} \Delta M^2_\beta$  \hfill (7)} \\
 trilinear  couplings &:  $A_k = d_k A_0   + d'_k M_{1/2}$         \hfill
(8)
\end{tabular}
\refstepcounter{equation}
\refstepcounter{equation}
\label{eq:gaugino}
\refstepcounter{equation}
\label{eq:squark}
\refstepcounter{equation}
The index $i$ runs over the gauge groups $i=SU(3)$, $SU(2)$, $U(1)$.
To this order, the gauge couplings, and the gaugino and scalar mass
parameters of soft--supersymmetry breaking depend on the $Z$ transporters
\begin{eqnarray}
Z_i^{-1} =  1 + b_i \frac{\alpha_U}{ 4 \pi}
             \log\left(\frac{M_U}{ M_Z}\right)^2
\end{eqnarray}
with $b[SU_3, SU_2, U_1] = -3, \, 1, \, 33 / 5$;
the scalar mass parameters depend
also on the Yukawa couplings $h_t$, $h_b$, $h_\tau$
of the
top quark, bottom quark and $\tau$ lepton.
The coefficients $c_j$ for the slepton and squark doublets/singlets,
and for the two Higgs doublets, are linear combinations of the evolution
coefficients $Z$; the coefficients $c'_{j \beta}$ are of order unity.
The shifts $\Delta M^2_\beta$, depending implicitly on all the other
parameters, are nearly zero for the first two families of
sfermions but they can be rather large for the third family and for the
Higgs mass parameters. The coefficients $d_k$ of the trilinear
couplings $A_k$ [$k=t,b,\tau$] depend on the corresponding
Yukawa couplings and they are approximately unity for the
first two generations while being O($10^{-1}$)
and smaller if the Yukawa couplings are
large; the coefficients $d'_k$, depending on gauge
and Yukawa couplings, are of order unity.
Beyond the approximate solutions, the evolution equations
have been solved numerically in the present analysis  to
two--loop order \cite{R23} and the threshold effects have been
incorporated at the low scale in one-loop order \cite{R24}.
Solutions of the renormalization group equations have been
obtained meanwhile up to three-loop order \cite{R25}, matching
the new two-loop order results of the threshold corrections
\cite{R26}. The 2-loop effects as given in
Ref.~\cite{R27} have been included for
the neutral Higgs bosons and the $\mu$ parameter.

\subsection{Gauge Coupling Unification}

\begin{figure*}[tb]
\setlength{\unitlength}{1mm}
a) \hspace{8cm} b) \\
\rule{0mm}{0mm} \hspace{.5cm}
\epsfig{figure=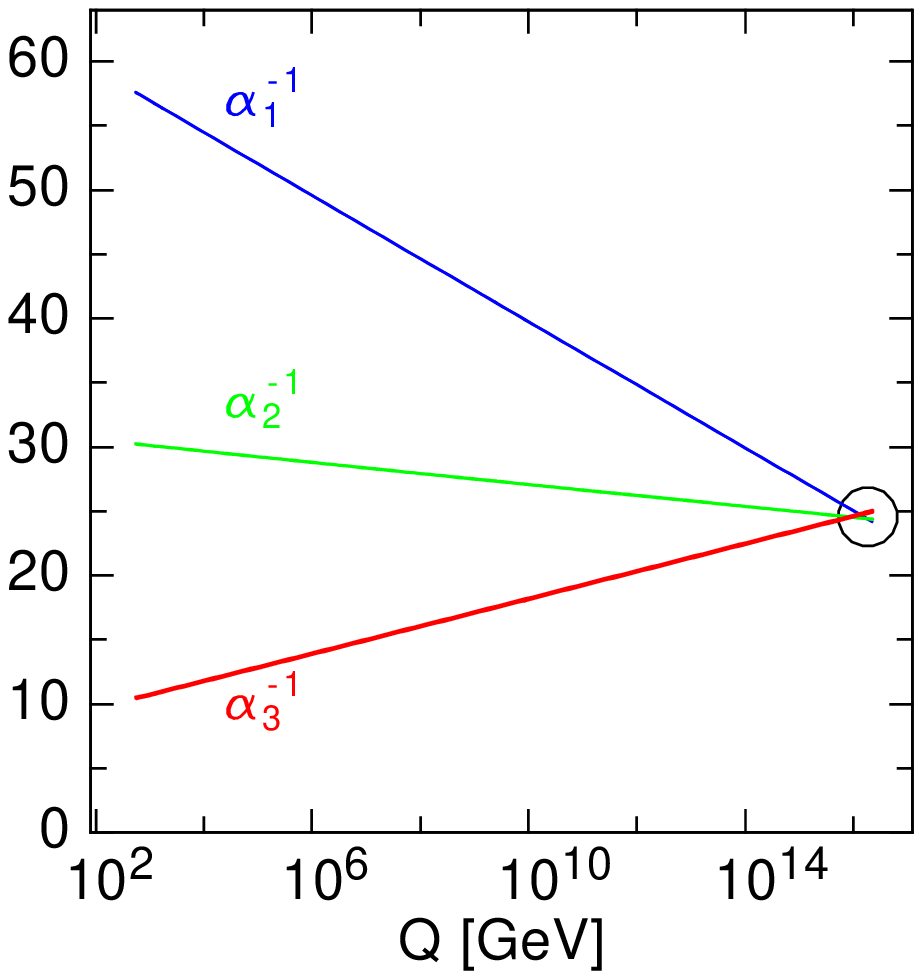, width=6cm, bb=100 380 360 650}\hspace{2cm} 
\epsfig{figure=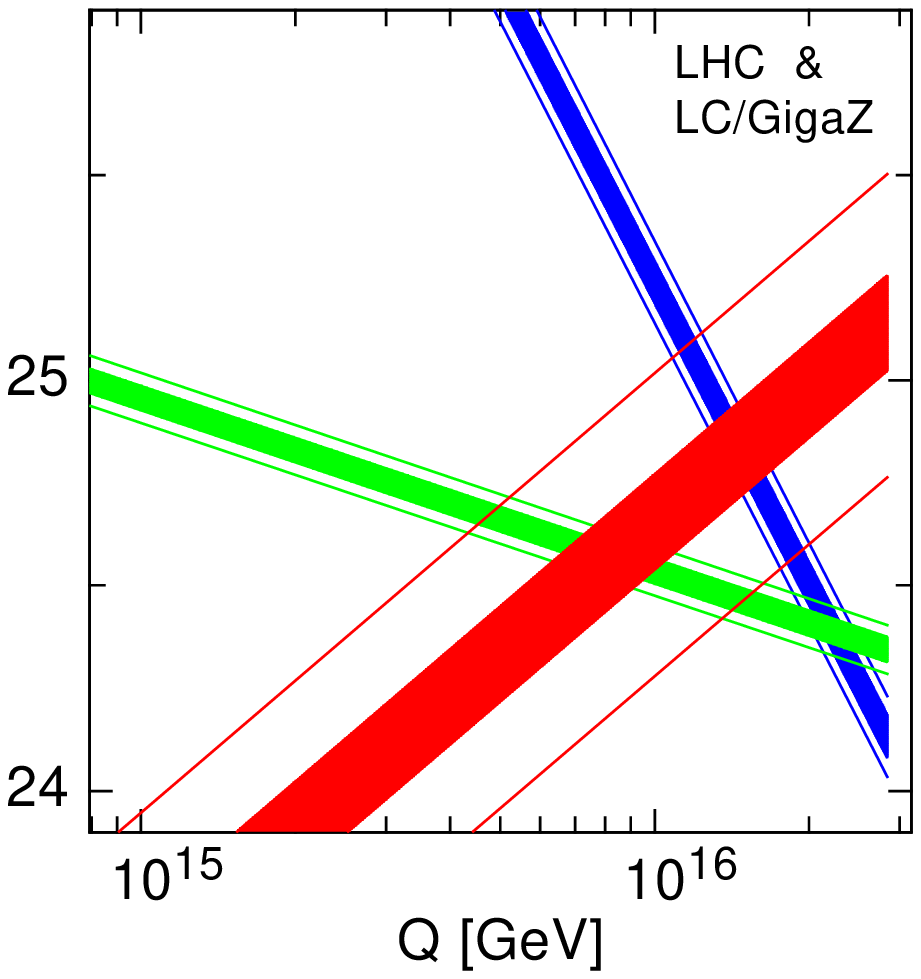, width=6cm, bb=100 380 360 650}
\vspace{-2em}
\caption{(a) Running of the inverse gauge couplings. (b) Close-up of
the unification region. The thin lines represent uncertainties 
based on present data, 
the solid areas demonstrate the improvement expected by future GigaZ
             analyses.}
\label{fig:gauge}
\end{figure*}

Measurements of the gauge couplings at the electroweak scale
support very strongly the unification of the couplings at a scale
$M_U \simeq 2\times 10^{16}$~GeV \cite{R2}.
The precision, at the per--cent level, is
surprisingly high after extrapolations over
fourteen orders of magnitude in the energy
from the electroweak scale to the grand unification scale $M_U$.
Conversely, the
electroweak mixing angle has been predicted in this approach at the
per--mille level. The evolution of the gauge couplings from
low energies to the GUT scale $M_U$ has been carried out at two--loop
accuracy
in the $\overline{DR}$ scheme.
The gauge couplings do not meet exactly, cf. Fig.~\ref{fig:gauge} and
Tab.~\ref{tab:gauge}.
The differences are to be attributed to high-threshold effects
at the unification
scale $M_U$ and the quantitative evolution implies
important constraints on the particle content at $M_U$
\cite{R28}. 

\begin{table*}[tb]
\centering
\begin{tabular}{|c||c|c|}
\hline
 & Present/''LHC'' & GigaZ/''LHC+LC'' \\
\hline \hline
$M_U$ & $(2.36 \pm 0.06)\cdot 10^{16} \, \rm {GeV}$ &
           $ (2.360 \pm 0.016) \cdot 10^{16} \, \rm {GeV}$ \\
$\alpha_U^{-1}$ & $  24.19 \pm 0.10 $ &  $ 24.19 \pm 0.05 $\\ \hline
$\alpha_3^{-1} - \alpha_U^{-1}$ & $0.97 \pm 0.45$ & $0.95 \pm 0.12$ \\
\hline
\end{tabular}
\caption{Precision of extraction of the unified gauge coupling $\alpha_U$, derived
from the meeting point of $\alpha_1$ with $\alpha_2$, and the strong coupling
$\alpha_3$ at the GUT scale $M_U$. The columns demonstrate the results for the 
expected precision from LEP and LHC data, as well as the improvement due to a
GigaZ linear collider analysis.}
\label{tab:gauge}
\end{table*}

\subsection{Gaugino and Scalar Mass Parameters}

\begin{figure*}
\setlength{\unitlength}{1mm}
\begin{center}
\begin{picture}(160,74)
\put(4,-81){\mbox{\epsfig{figure=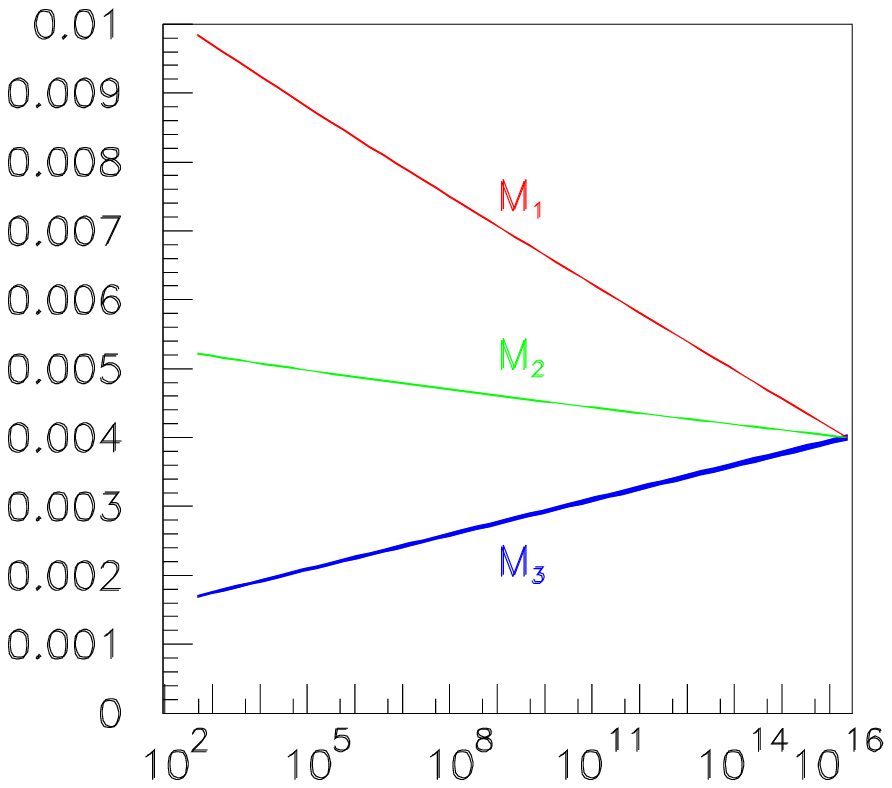,width=16cm}}}
\put(83,-81){\mbox{\epsfig{figure=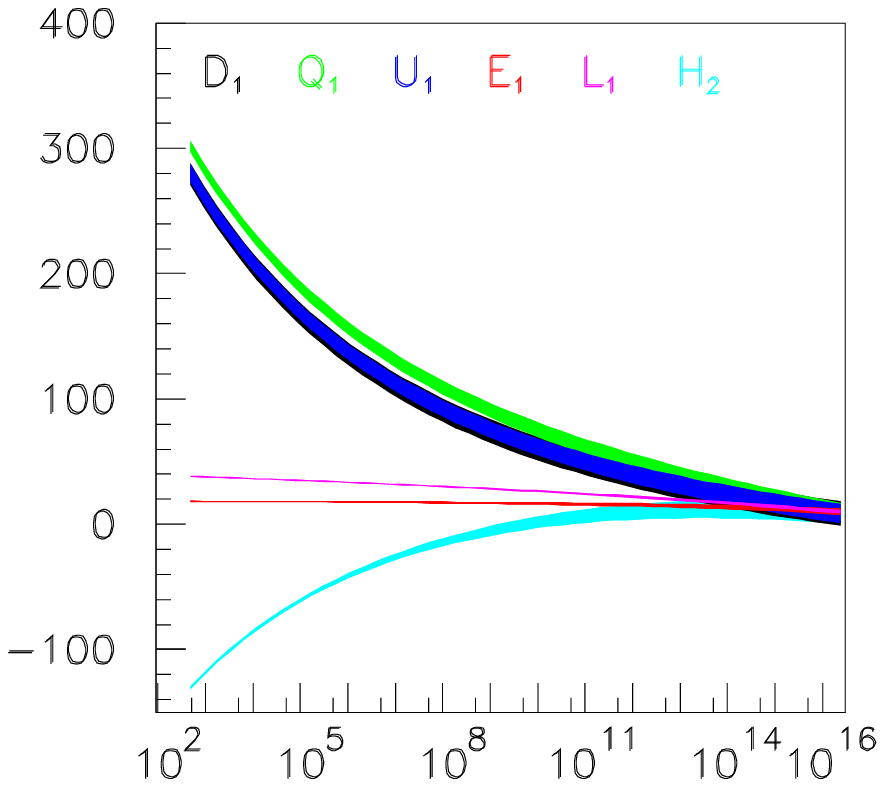,width=16cm}}}
\put(-1,70){\mbox{\bf (a)}}
\put(14,68){\mbox{$1/M_i$~[GeV$^{-1}$]}}
\put(57,-3){\mbox{$Q$~[GeV]}}
\put(80,70){\mbox{\bf (b)}}
\put(95,68){\mbox{$M^2_{\tilde j}$~[$10^3$ GeV$^2$]}}
\put(145,-3){\mbox{$Q$~[GeV]}}
\end{picture}
\end{center}
\vspace{-2em}
\caption{Evolution, from low to high scales, of
(a) gaugino mass parameters and
 (b) first-generation sfermion mass parameters and
     the Higgs mass parameter $M^2_{H_2}$.}
\label{fig:sugra_LHC}
\end{figure*}

The results for the evolution of the mass parameters from the electroweak
scale to the GUT
scale $M_U$ are shown in Fig.~\ref{fig:sugra_LHC}.
On the left of Fig.~\ref{fig:sugra_LHC}
the evolution is presented for the
gaugino parameters $M^{-1}_i$. It clearly
is under excellent control for the
model-independent reconstruction
of the parameters and the test of universality
in the $SU(3) \times SU(2) \times U(1)$ group space.
In the same way the evolution of the scalar mass parameters can be
studied, presented in Fig.~\ref{fig:sugra_LHC}~(b)
for the first/second generation.
While the slepton parameters can be determined very accurately,
the accuracy deteriorates for the squark parameters
and the Higgs parameter $M^2_{H_2}$.  

\section{Left-right symmetric extension}

The complex structure observed in the neutrino sector
has interesting consequences for the properties of the sneutrinos,
the scalar supersymmetric partners of the neutrinos. These novel elements
require the extension of the minimal supersymmetric
Standard Model MSSM, e.g., by a superfield including the
right-handed neutrino field and its scalar partner \cite{R7}.
If the small neutrino masses are generated
by the seesaw mechanism \cite{R8},
a similar type of spectrum is induced in the scalar sector, splitting
into light TeV-scale and very heavy masses.
The intermediate seesaw scales will affect
the evolution of the soft
mass terms which break the supersymmetry
at the high (GUT) scale,
particularly in the third generation
with large Yukawa couplings \cite{R3,R10}.
This will provide the opportunity to measure, indirectly, the
intermediate seesaw scale of the third generation.  

\vspace{1.5ex}
If sneutrinos are lighter than charginos and the second lightest
neutralino, as encoded in SPS1a$'$, they decay only to final states $\snl
\to \nu_l \, \neu_1$ that are invisible and pair-production is useless
for studying these particles.
However, in this configuration sneutrino masses
can be measured in chargino decays to sneutrinos and leptons \cite{R10}:
\begin{equation}
\cha^\pm_1 \to l^\pm \, \sn_l^{(*)},
\label{eq:chadec}
\end{equation}
with the charginos pair-produced in $e^+e^-$ annihilation. These
two-particle
decays develop sharp edges at the endpoints of the lepton energy
spectra.
Sneutrinos of all three generations can be explored this way.
The errors for the first and second generation sneutrinos are expected at
the level of 400 MeV, doubling for the more involved analysis of the third
generation. 

\vspace{1.5ex}
The measurement of the seesaw scale will be illustrated
in an SO(10) model in which the matter superfields of the three
generations belong to
16-dimensional representations of SO(10) and the standard Higgs
superfields
to 10-dimensional representations while a Higgs superfield in the
126-dimensional representation generates Majorana masses for
the right-handed neutrinos. As a result, the Yukawa couplings
in the neutrino sector
are proportional to the up-type quark mass matrix, for which
the standard texture is assumed. The SO(10) symmetry is
broken to the Standard Model SU(3)$\times$SU(2)$\times$U(1)
symmetry at the grand unification scale $M_U$ directly.
For simplicity the soft masses
in the Higgs sector will be identified with the matter sector.

Assuming that the Yukawa couplings are the same for up-type quarks
and neutrinos at the GUT scale and that the Majorana mass matrix of
the right-handed neutrinos has a similar structure, one obtains
a (weakly) hierarchical neutrino mass spectrum and nearly bi-maximal
mixing
for the left-handed neutrinos  \cite{R29}.
In this class of models the masses of the right-handed neutrinos
are also hierarchical, very roughly $\propto m^2_{\rm up}$, and
the mass of the heaviest neutrino is given by
$M_{\RR_3} \sim m^2_t / m_{\nu_3}$.
For $m_{\nu_3} \sim 5 \times 10^{-2}$ eV, the heavy neutrino mass of the
third generation amounts to $\sim 6 \times 10^{14}$ GeV, {\it i.e.} a
value
close to the grand unification scale $M_U$.  

\begin{figure}
\setlength{\unitlength}{1mm}
\begin{picture}(80,90)
\put(0,17){\mbox{\epsfig{figure=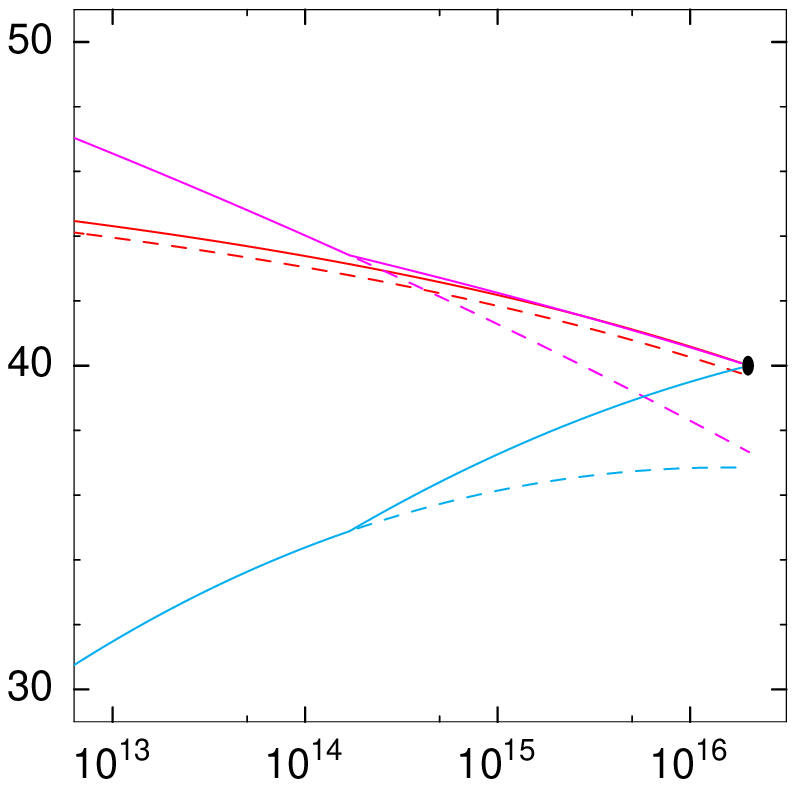,height=6.6cm,width=6.9cm}}}
\put(6,86){\mbox{$M^2_{\tilde j}$~[$10^3$ GeV$^2$]}}
\put(58,13){\mbox{$Q$~[GeV]}}
\put(8,73){\mbox{\small $M^2_{L_3}$}}
\put(9,59){\mbox{\small $M^2_{E_3}$}}
\put(10,36){\mbox{\small $M^2_{H_2}$}}
\put(36.5,30){\mbox{\small $\nu_{R_3}$}}
\put(32.3,30){\vector(0,1){5}}
\put(31,75){\mbox{\small ----- $\nu_R$, $\tilde{\nu}_R$ included}}
\put(31,70){\mbox{\small - - -  $\nu_R$, $\tilde{\nu}_R$ excluded}}
\end{picture}
\vspace{-7em}
\caption{Evolution of third generation slepton mass parameters and
     Higgs mass parameters  $M^2_{H_2}$ in LR-SUGRA.}
\label{fig:sugraRnu}
\end{figure}
Since the $\nu_R$ of the third generation is unfrozen only beyond
$Q=M_{\nu_R}$ the impact of the LR extension becomes visible in the
evolution of the scalar mass parameters
only at very high scales.
In Fig.~\ref{fig:sugraRnu} the evolution of $M^2_{\tilde E_3}$,
$M^2_{\tilde L_3}$ and $M^2_{H_2}$ is displayed
for illustrative purposes.  The full lines
include the effects of the right--handed neutrino, which should
be compared with the dashed
lines where the $\nu_R$ effects are removed.
The scalar mass parameter $M^2_{\tilde E_3}$ appears unaffected by the
right--handed sector, while $M^2_{\tilde L_3}$ and $M^2_{H_2}$ clearly
are.
Only the picture including $\nu_R$, $\tilde{\nu_R}$ is compatible
with the unification assumption.
The kinks in the
evolution of $M^2_{{\tilde L}_3}$ and $M^2_{\tilde {H_2}}$ can be traced
back
 to the
fact that around $10^{14}$ GeV the third generation (s)neutrinos
become quantum mechanically effective, given a large enough neutrino
Yukawa coupling
to influence the evolution of these mass parameters. 

To leading order, the solutions of the renormalization group
equations for the masses of the scalar selectrons and the $e$-sneutrino can be
expressed by
the high scale parameters $m_{16}$ and $M_{1/2}$, and the D-terms.
Analogous representations can be derived, to leading order, for the scalar
masses of the third generation, complemented however by additional
contributions $\Delta_\tau$
and $\Delta_{\nu_\tau}$ from the standard tau Yukawa term and the
Yukawa term in the tau neutrino sector, respectively:
\begin{align}
\label{eq:mtaur}
 m^2_{\tilde{\tau}_R} &=  m^2_{\tilde{e}_R}
                      - 2 \Delta_\tau          + m^2_\tau
\\
 m^2_{\tilde{\tau}_L} &=  m^2_{\tilde{e}_L}
                      - \Delta_\tau - \Delta_{\nu_\tau} + m^2_\tau,
\\
 m^2_{\tilde{\nu}_{\tau L}} &=  m^2_{\tilde{\nu}_{e L}}
                       - \Delta_\tau - \Delta_{\nu_\tau}.
\label{eq:mtausnu}
\end{align}
The contribution $\Delta_{\nu_\tau} = \Delta_{\nu_\tau}[M_\RR]$ 
carries the information on the value
of the heavy right-handed neutrino mass.   \\

\begin{figure*}
\centering
\epsfig{figure=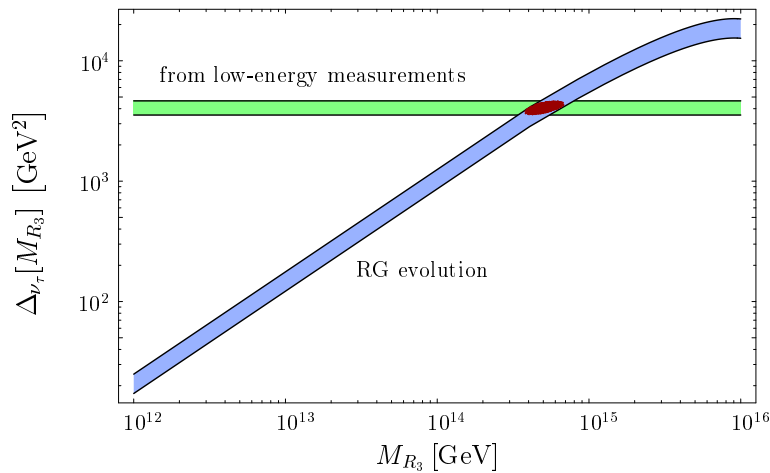, width=4in, bb=40 441 435 681, clip=true}
\vspace{-3em}
\caption{Shift $\Delta_{\nu_\tau}$
in the evolution of the tau-neutrino mass as calculated
from the renormalization group equations,
eq.~\eqref{eq:Dnuev} (blue band) and compared with
low-energy mass measurements, eq.~\eqref{eq:deltaM3} (green band).
The widths of the bands indicate estimated
one-standard-deviations errors of the experimental input parameters. The red
crossing region is the statistical combination which determines the
neutrino seesaw scale
$M_{\RR_3}$ of the third generation.}
\label{fig:MR}
\end{figure*}

The effect of the right-handed neutrinos can be identified by
evaluating the sum rule for the $\Delta_{\nu_\tau}$ parameter:
\begin{equation}
\begin{aligned}
2 \Delta_{\nu_\tau}[M_{\RR_3}] &= (3 m^2_{\tilde{\nu}_{e \LL}}
                - m^2_{\tilde{e}_\LL} - m^2_{\tilde{e}_\RR}) \\&\,-
                  (3 m^2_{\tilde{\nu}_{\tau \LL}} -
                  m^2_{\tilde{\tau}_1} - m^2_{\tilde{\tau}_2}) - 2
m_\tau^2.
\end{aligned}
\label{eq:deltaM3}
\end{equation}
This relation holds exactly at tree-level and gets modified
slightly only by small corrections at the one-loop level.
The particular form of eq.~(\ref{eq:deltaM3})
implies that the effects of the $\tau$ Yukawa coupling cancel.
It follows from the renormalization group equations
that $\Delta_{\nu_\tau}[M_{\RR_3}]$ is of the order
$Y^2_\nu \log{M^2_{\rm GUT}/M^2_{\RR_3}}$.
Since the Yukawa coupling $Y_\nu$ can be
estimated in the seesaw mechanism by the mass
$m_{\nu_3}$ of the third light neutrino,
$Y_\nu^2 = m_{\nu_3} M_{\RR_3} / (v \, \sin\beta)^2$,
the parameter $\Delta_{\nu_\tau}[M_{R_3}]$ depends approximately linearly
on the mass $M_{\RR_3}$:
\begin{equation}
\begin{aligned}
\Delta_{\nu_\tau}[M_{R_3}] \simeq  &\,\frac{m_{\nu_3} M_{\RR_3}}
                                         {16 \pi^2 (v \, \sin\beta)^2} \\[-1ex]
&\times \left( 3 m^2_{16}  + A^2_0 \right)
\log\frac{M^2_{\rm GUT}}{M^2_{\RR_3}},
\end{aligned}
\label{eq:Dnuev}
\end{equation}
so that it can be determined very well.
Inserting the value for $m_{16}$, pre-determined in the charged slepton
sector, and the trilinear coupling $A_0$ from stop mixing
\cite{R19},
$M_{\RR_3}$, cf.~Fig.~\ref{fig:MR}, can finally be calculated.
Assuming hierarchical neutrino masses, one obtains
\begin{equation}
  M_{\RR_3} = 3.7 \; \dots \; 6.9 \times 10^{14} \gev,
\end{equation}
to be compared with the initial value $M_{\RR_3} = 6 \times 10^{14}$ GeV.
This analysis thus provides us with a unique estimate of the
high-scale $\nu_R$ mass parameter $M_{\RR_3}$.   \\

\section{String Effective Theories}

In the supergravity models analyzed above the supersymmetry
breaking mechanism in the
hidden sector is shielded from the eigen--world. Four--dimensional
strings however give rise to a minimal set of fields for inducing
supersymmetry breaking, the dilaton $S$ and the moduli $T$
superfields which are generically present in large classes of
4--dimensional heterotic string theories \cite{R11}. The vacuum
expectation
values of $S$ and $T$, generated by genuinely non--perturbative
effects, determine the soft supersymmetry breaking parameters. 

The properties of the supersymmetric theories
are quite different for dilaton and moduli dominated scenarios. This
can be quantified by introducing a mixing angle $\theta$,
characterizing the $\tilde S$ and $\tilde T$ wave functions of the
Goldstino, which is associated with the breaking of supersymmetry and
which is absorbed to generate the mass of the gravitino:
$\tilde G = \sin \theta \, \tilde{S} + \cos \theta \, \tilde{T}$.
The mass
scale is set by the second parameter of the theory, the gravitino mass
$m_{3/2}$.  

A dilaton dominated scenario, i.e.~$\sin \theta \to 1$, leads to
universal boundary conditions of the soft supersymmetry breaking
parameters.  Universality is broken only slightly by small loop effects.
On the other hand, in moduli dominated scenarios,
$\cos\theta \to 1$, the gaugino mass parameters are universal to lowest
order, but universality is not realized
for the scalar mass parameters. The breaking is characterized
by modular weights $n_j$ which quantify the couplings between the
matter and the moduli fields in orbifold compactifications.
Within one generation significant
differences between left and right field components and between
sleptons and squarks can occur. 

In leading order the masses \cite{R30}
are given by the following expressions for
the gaugino sector,
\begin{eqnarray}
M_i &=&  - g_i^2 m_{3/2} s {\sqrt{3} \sin \theta} + ...
\end{eqnarray}
and for the scalar sector,
\begin{eqnarray}
M_{\tilde j}^2 &=& m^2_{3/2} \left(
  1 + n_j \cos^2 \theta \right) + ...
\end{eqnarray}
A mixed dilaton/moduli superstring scenario has been analyzed in detail
with dominating dilaton
component, $\sin^2 \theta = 0.9$, and with different couplings of
the moduli field to the (L,R) sleptons, the (L,R) squarks and to the Higgs
fields, corresponding to O--I representation $n_{L_i} = -3$, $n_{E_i} =
-1$,
$n_{H_1} =n_{H_2}=-1$,
 $n_{Q_i} = 0$, $n_{D_i} = 1$ and $n_{U_i} = -2$,
an assignment that is adopted quite frequently in the literature.
The gravitino mass is chosen to be 180~GeV in this analysis. 

\begin{figure*}
\rule{0mm}{0mm}\\[-3em]
\epsfig{figure=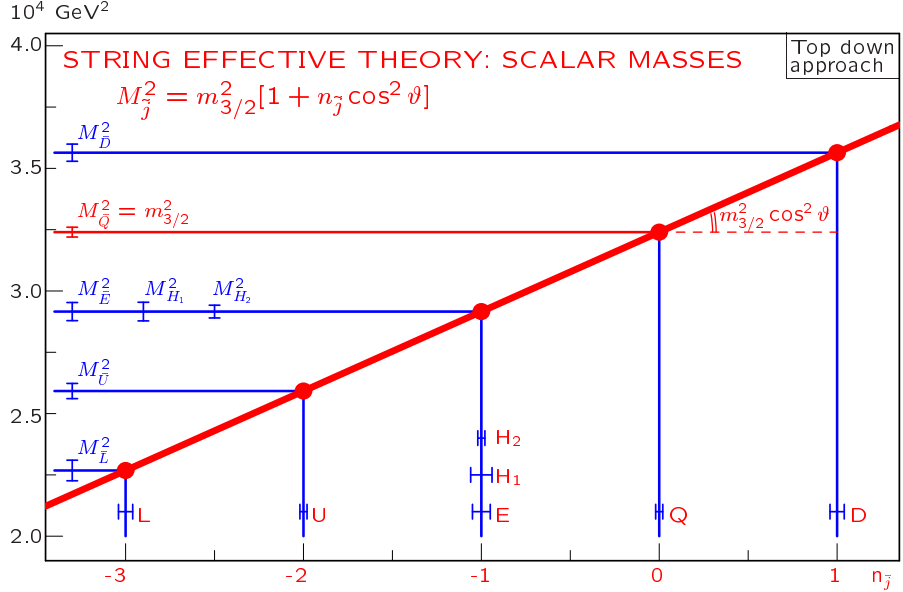, height=14cm, angle=270}
\vspace{-5.5em}
\caption{Extraction of parameters of
superstring inspired supersymmetry breaking scenario from gaugino and scalar
evolution.}
\label{fig:chewfraut}
\end{figure*}
Given this set of superstring induced parameters, the evolution of the
gaugino and scalar mass parameters can be exploited to determine the
modular weights $n$. The result is shown in Fig.~\ref{fig:chewfraut} which
demonstrates
quite nicely how stringently this theory can be tested by analyzing
the integer character of the entire set of weights. 

Thus, high-precision measurements at high energy proton and $e^+e^-$
linear colliders provide access to crucial derivative parameters
in string theories.

\section{Conclusions}

In supersymmetric theories stable extrapolations can be
performed, by renormalization group techniques,
from the electroweak scale to the grand unification
scale close to the Planck scale.
Such extrapolations are made possible by high-precision
measurements of the low-energy parameters.
In the near future an enormous {\it corpus} of information
is expected to become available if measurements at LHC and
prospective $e^+e^-$ linear colliders are combined to draw
a comprehensive and high-precision picture of supersymmetric
particles and their interactions. 

Supersymmetric theories and their breaking mechanisms have
simple structures at high scales.
Extrapolations to these scales are therefore crucial to reveal
the fundamental supersymmetric theory, including its symmetries
and its parameters. The extrapolations can thus
be used to explore physics phenomena at a scale where,
eventually, particle physics is linked to gravity. 

We have reviewed three interesting scenarios in this context.
The universality of gaugino and scalar mass parameters
in minimal supergravity can be demonstrated very clearly. Intermediate
scales, like seesaw scales in left-right symmetric theories,
affect the evolution of the scalar mass parameters. 
Their effect on the
mass parameters of the third generation can be exploited to
determine the seesaw scale. Finally it has been shown that integer
modular weights can be measured accurately in string effective
theories, scrutinizing such approaches quite stringently. 

\vspace{1.5ex}
Many more refinements of the theoretical calculations and future
experimental analyses will be necessary to expand the pictures we
have described in this review. Prospects of exploring
elements of the ultimate unification of the interactions provide
a strong impetus to this endeavor. 



\end{document}